\documentclass[reprint,tikz,prb,twocolumn,superscriptaddress]{revtex4-1}

\usepackage{amsmath,amssymb}
\usepackage{graphicx}

\usepackage{float}
\usepackage[usenames,dvipsnames]{xcolor}
\usepackage{amsthm,comment}
\usepackage{booktabs}
\usepackage[export]{adjustbox}
\usepackage[section]{placeins}
\usepackage[caption=false]{subfig}
\usepackage[percent]{overpic}
\bibpunct{[}{]}{,}{n}{}{}
\definecolor{red1}{HTML}{FF4136}
\definecolor{green1}{HTML}{00802b}

\usepackage[hidelinks]{hyperref} %boxes hidden, remove hidelinks if boxes are desired.

\hypersetup{colorlinks=true,citecolor=Red,linkcolor=Blue,urlcolor=Blue}
\usepackage[export]{adjustbox}

\newcommand{\<}{\langle}
\renewcommand{\>}{\rangle}

\graphicspath{{Figures/}}

\begin{document}

\title{Properties of La$_{0.7}$Ca$_{0.3}$MnO$_3$ Under Extreme Tensile Strain}

\author{C. \c{S}en}
\email{Corresponding author. Email: csen@lamar.edu}
\affiliation{Department of Physics, Lamar University, Beaumont, Texas, 77710, USA}

\author{E. Dagotto}
\affiliation{Department of Physics and Astronomy, The University of Tennessee, Knoxville, Tennessee 37996, USA}
\affiliation{Materials Science and Technology Division, Oak Ridge National Laboratory, Oak Ridge, Tennessee 37831, USA}

\begin{abstract}
The complex phase diagram of manganites with simultaneously active spin, charge, orbital, and lattice degrees of freedom continues providing surprises.
In a recent groundbreaking experiment, membranes of the perovskite manganite La$_{0.7}$Ca$_{0.3}$MnO$_3$ (LCMO) deposited on a flexible
polymer layer were strained up to 8\% [S.~S. Hong {\it et al.}, Science {\bf 368}, 71 (2020)], 
much more than achieved by regular strain induced by a rigid substrate. 
By increasing this strain a metal-insulator transition was reported. Here we reproduce
the results of the experiments using Monte Carlo simulations of the two-orbital double-exchange model including Jahn-Teller distortions, at hole density $x=1/3$. 
The full phase diagram with varying temperature and Jahn-Teller coupling $\lambda$ is presented. 
When the bandwidth $W$ of mobile electrons is reduced, thus when the effective Jahn-Teller coupling $\lambda/W$ is increased,
a metal-insulator transition is found in our simulations, between a ferromagnetic metallic state with uniform charge distribution 
and an insulator with diagonal charge
stripes that retains its ferromagnetic character. In between the hole-rich diagonals, staggered orbital order occurs. 
We also report resistivity and magnetization measurements alongside with spin correlations and charge structure factors.
Our overall conclusions are in agreement with the recent experimental and density functional theory 
results by Hong {\it et al.}, and we confirm much earlier ground state predictions of striped ferromagnetic order 
using energy optimization techniques by T. Hotta {\it et al.}, Phys. Rev. Lett. {\bf 86}, 4922 (2001). 
The experimental observation of one of the states predicted by theory suggests that diagonal stripes
could be achieved at other hole densities as well, such as $x=1/4$, if LCMO membranes with that hole doping were subject to similar strains. 
 
\end{abstract}

\maketitle

%%%% ================================================================= %%%%
%%%% ================================================================= %%%%
\section{Introduction} 

%\subsection{Historical perspective of manganites}

Manganites are among the most studied transition metal oxides because of their fascinating
colossal magnetoresistance (CMR) effect, where relatively small magnetic fields can trigger changes
in the resistivity by several orders of magnitude~\cite{ramesh,deteresa,moreo,dagotto,salamon}. 
In addition to this remarkable nonlinear response, manganites are well known for at least three 
additional reasons: 

({\it i}) They present a rich phase diagram that includes the metallic
ferromagnetic (FM) phase induced by ``double exchange'' where itinerant $e_g$ electrons move
in a background of localized $t_{2g}$ spins. Moreover, a wide variety of charge-ordered phases has
been unveiled for narrow bandwidth manganites~\cite{dagotto}, such as 
La$_{1-x}$Ca$_{x}$MnO$_3$ (LCMO) near $x=0.5$ where a complex pattern of spin, orbital, charge, 
and Jahn-Teller lattice distortions emerge creating zigzag chains, among other 
self-organized textures. Several other charge-ordered (CO) patterns have been proposed 
and some were confirmed experimentally, as discussed below. 

({\it ii}) Manganites present the exotic effect of ``electronic phase separation'' (EPS) induced
by the close similarity in energy of vastly different states~\cite{dagotto,sen2004,cengiz1,cengiz2,tomioka,LSMO1}, 
both metallic and insulators, causing the CMR effect. This leads to
an intrinsic frustration, not obvious at the Hamiltonian level. EPS produces both 
novel complex states as well as nano-meter scale separation of phases 
that usually manifests via the coexistence of features of competing states in the same sample. 
For extreme cases such as (La$_{1-y}$Pr$_{y}$)$_{1-x}$Ca$_{x}$MnO$_3$ (LPCMO) the phase-separated scale reaches the sub-micrometer scale~\cite{cheong}.
This mixture of properties creates nonlinear responses under external perturbations~\cite{tokura,tokura2,ichikawa,argyriou,adams2000}. 
Such an unusual property could be useful in devices~\cite{shen1,mitchell,shen2,shen3,shen4,EPS1,EPS2}. 
Percolation has previously been shown to be common in this context and it is at the heart of the
CMR in magnetic fields~\cite{dagotto,lozanne,percolative1,percolative2}. In general, a high level of ``complexity'' exists in materials with strong phase
competition as in these manganese oxides~\cite{complex}.

({\it iii}) In addition, manganites are used extensively in preparing artificial superlattices of
oxides. In this context, La$_{1-x}$Sr$_{x}$MnO$_3$ (LSMO) is primarily employed due to its excellent metallic properties.
However, exploiting the unusual charge-ordered properties of LCMO in the form of superlattices seems another natural path to follow.
For instance, La$_{2/3}$Ca$_{1/3}$MnO$_3$ has been used in conjunction with a high critical temperature superconductor 
YBa$_2$Cu$_3$O$_7$ to unveil exotic properties of their heterostructures, such as a change in the orbital occupation 
induced in the cuprate's side of the interface~\cite{keimer1,keimer2,salafranca}.
But the LCMO avenue for superlattices has been addressed at a relatively much lower pace, as compared with the far more extensive
use of LSMO.

From the theory perspective, a variety of exotic charge, spin, and orbital states have been reported in
various regions of parameter space, particularly for narrow bandwidth manganites with robust Jahn-Teller coupling,
namely with LCMO as a natural candidate for its realization~\cite{hotta2,ye2009,dong2009,liang2011,sen2012,sinclair}. 
Of particular importance in our efforts is the early prediction of ferromagnetic charge-ordered stripes 
by Hotta {\it et al.}~\cite{hotta2001}, as explained below.

\subsection{Recent experiments and earlier theory}

Recent ground-breaking experimental results~\cite{hong2020} provided new impetus to research in the rich phase diagram of LCMO. It was reported in~\cite{hong2020} that extreme tensile (increase in the lattice constants) strain was achieved in LCMO membranes, exceeding 8\% uniaxially, a strain never reported before. This was made possible by growing heterostructures of SrCa$_2$Al$_2$O$_6$, SrTiO$_3$, and LCMO. The SCAO component was used because it nearly perfectly matches the lattice spacings of LCMO. Moreover, SCAO was ``sacrificed'' after the growth by mere dissolution in water. By this procedure, the top oxide film was detached from the substrate and transferred onto a polymer, which is then used as a flexible substrate. The four sides of the polymer were stretched with micromanipulators and the strong adhesion polymer-oxide membrane provides strain transfer to LCMO to unprecedented levels up to 0.4~nm, far beyond typical strains in crystalline superstructures.

%-----------------------------------------------------------------------------------------
\begin{figure}[H]
\begin{center}
% trim={<left> <lower> <right> <upper>}
 \begin{overpic}[trim = 0cm 0.0cm 0mm 0mm,angle=270,
 width=0.44\textwidth]{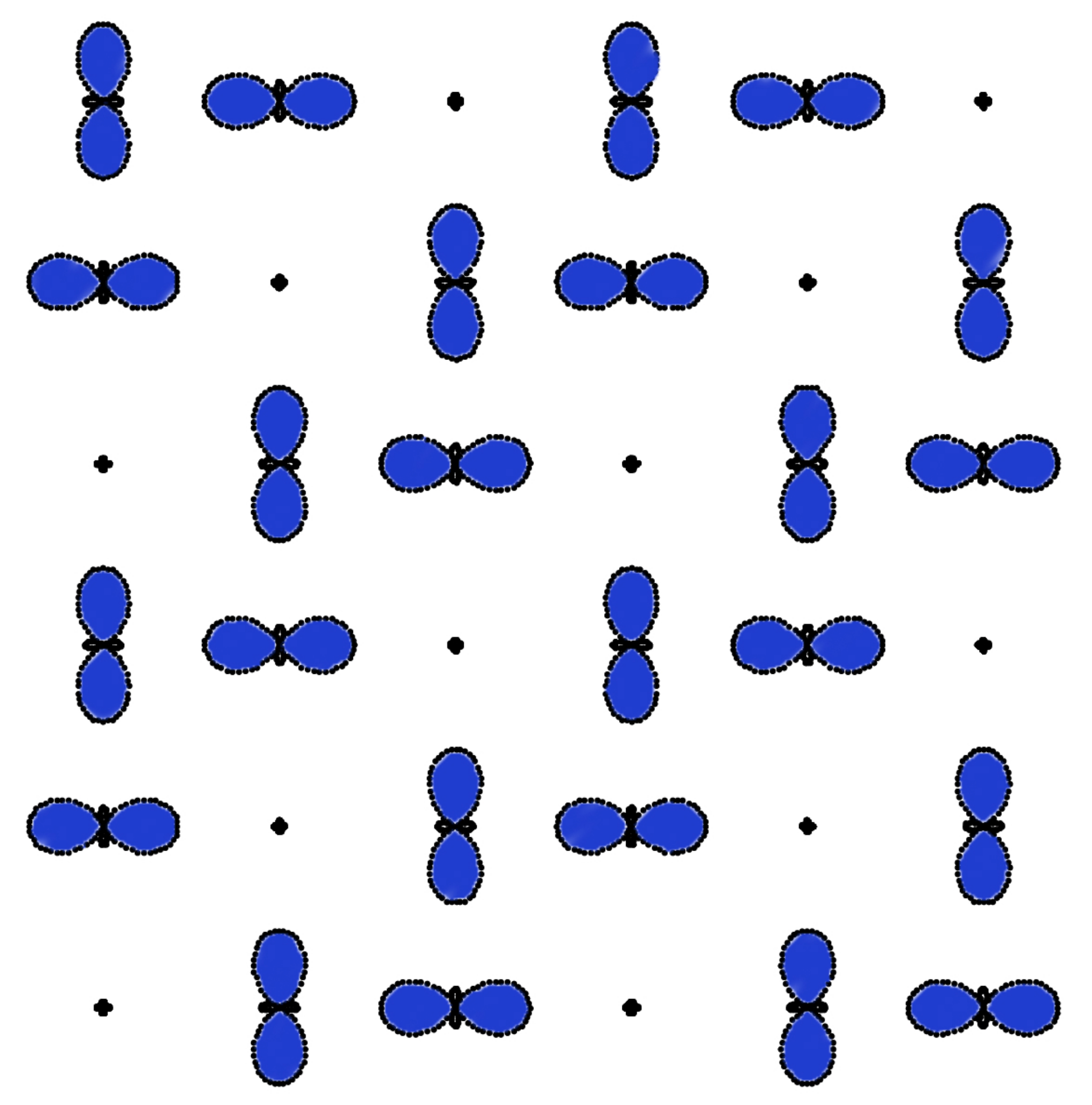}
\end{overpic}
\end{center}
\vspace{-0.6cm}
\caption{Idealized diagonal stripe states predicted by Hotta {\it et al.}~\cite{hotta2001} and recently observed in
LCMO membranes experiments by Hong {\it et al.}~\cite{hong2020}. Shown are only the orbital degrees of freedom, where the charge is located. The overall hole
density is $x=1/3$. Stripes running from top left to bottom right are clearly visible. The spin is ferromagnetic, not shown.}
\label{fig0}
\end{figure}
%-----------------------------------------------------------------------------------------

The transport properties of the LCMO membrane were shown to evolve together with the strain~\cite{hong2020}. By using electrodes near the center of the membrane, thus avoiding edge effects, the resistivity vs temperature was measured. As the strain increased, the double-exchange induced metallic state with a critical Curie temperature of $T_C \sim 250$~K was gradually suppressed and eventually replaced by an insulating state at low temperatures. The applied tensile strain naturally reduces the mobility of electrons via the reduction of the Mn-Mn hopping amplitudes, thus indirectly enhancing the effects of correlations
via the reduction of the bandwidth $W$. This is especially important with regards to the Jahn-Teller electron-lattice coupling, known to be crucial in manganites. Employing Density Function Theory (DFT) techniques, the insulating state found experimentally at high strain was proposed to be an in-plane magnetically ferromagnetic state although in an overall A-AFM stacking configuration~\cite{hong2020}. 
At high strain, the active Jahn-Teller modes provide a peculiar ordered distribution of charge, with non-equivalent Mn$^{4+}$ and Mn$^{3+}$ sites, resembling {\it diagonal stripes} from the perspective of the Mn lattice. This state had not been observed before experimentally to our knowledge.

The recently reported DFT calculation results~\cite{hong2020} are compatible with the highly strained LCMO experimental results. However, the original prediction of such an exotic state, involving $x=1/3$ doped Mn oxide planes with spin ferromagnetic order and a diagonal striped arrangement of charge, was formulated long ago in the context of double exchange models for manganites, precisely in the regime of large Jahn-Teller coupling. More specifically, in Ref.~\onlinecite{hotta2001}, the zero temperature stabilization of two-dimensional ferromagnetic states with stripe order at the particular hole densities $x=1/m$ ($m$=integer) was predicted. For $m=2$, the charge pattern resembled the famous CE state of manganites~\cite{dagotto} with regards to lattice, charge, and orbital degrees of freedom, albeit with ferromagnetic order. To obtain those results the Mn-Mn $t_{2g}$ superexchange among the localized spins was considered to be weak. For $m=3$, the case of relevance in~\cite{hong2020} and in the present publication, the state was novel at the time and was not observed experimentally until now. For $m=4$, an equally exotic order is anticipated and future experiments using highly strained LCMO at $x=0.25$ may confirm the existence of such a state. 

The early theory efforts~\cite{hotta2001} were at zero temperature, employing relaxation techniques to optimize the oxygen coordinates and, thus, exploiting the {\it cooperative} nature of the Jahn-Teller coupling. Coulombic interactions among electrons are effectively incorporated into the Jahn-Teller coupling, 
at least at the mean-field level~\cite{malvezzi}. 
Concomitantly with the FM order, staggered orbital order was also predicted in the $x=1/m$ stripes. This orbital degree of freedom acts as a ``pseudospin'' that establishes an analogy with the 
better-known stripes of nickelates and cuprates, where the orbital is not active but the spin background is staggered, namely antiferromagnetic. Increasing the Mn-Mn superexchange, a plethora of additional spin, charge, and orbital arrangements were also 
predicted~\cite{hotta2001}. 

In the present publication, the new exciting experimental results~\cite{hong2020} provide motivation to extend our previous zero temperature efforts in two directions. First, by employing Monte Carlo simulations we incorporate temperature effects and produce resistivity vs temperature curves that can be directly compared against experiments~\cite{hong2020}. This is a challenging theoretical task because of the multiplicity of metastable states that exist in the problem. Only a careful very slow  annealing process (costly in CPU time) from high to low temperatures allowed us to obtain reliable results that are in excellent agreement with experiments~\cite{hong2020}. In addition, by systematically increasing the value of the Jahn-Teller coupling $\lambda$, or equivalently, decreasing the value of the bandwidth $W$ by strain~\cite{hong2020}, we interpolate between a regime which is FM and metallic to another regime which is still FM but with exactly the charge stripes recently reported~\cite{hong2020}. This interpolation is achieved because the ratio $\lambda/W$ is the dimensionless parameter that matters to regulate the physics, thus reducing $W$ (tensile strain increases the lattice constants and
decreases the hopping amplitudes) amounts to modifying the strength of the Jahn-Teller coupling. Our effort then provides a unified rationale to explain the high strain LCMO results~\cite{hong2020} and brings to the forefront the early theoretical predictions that had not been confirmed until now~\cite{hotta2001}. Moreover, this success gives us extra confidence that  those early calculations~\cite{hotta2001} were reliable and allow us to predict 
a variety of other single oxide membrane experiments that could unveil  a variety of never observed before novel striped patterns.
In fact, a plethora of unusual states have been predicted over several years under various 
conditions~\cite{newstates1,newstates2,newstates3,newstates4,newstates5,newstates6} 
and the new procedure to achieve high strain is ideal to unveil those states.

The organization of this publication is as follows. In Sec.~II the model and many-body methodology is presented, emphasizing the annealing process
required for proper convergence. In Sec.~III, the main results are presented, summarized in a phase diagram. The numerical evidence justifying this phase
diagram is provided in this Section as well. Finally, in Sec.~IV brief conclusions are provided.

\section{Model and Method}

In our study, we use the standard double-exchange two-orbital lattice Hamiltonian for manganites 
defined on two-dimensional clusters~\cite{dagotto}. In the widely-employed limit of
infinite-Hund coupling, which considerably simplifies the Monte Carlo simulation, the model is:
\begin{widetext}
\begin{eqnarray}
  H &=& -\sum_{{\bf ia}\gamma \gamma'\sigma}
  t^{\bf a}_{\gamma \gamma'} 
  [\cos({\theta_{\bf i}/2})\cos({\theta_{\bf i+a}/2})
  +e^{-i(\phi_{\bf i}-\phi_{\bf i+a})}\sin({\theta_{\bf i}/2})\sin({\theta_{\bf i+a}/2)}]
  d_{{\bf i} \gamma \sigma}^{\dag}  d_{{\bf i+a} \gamma' \sigma}
%  -J_{\rm H} \sum_{\bf i}
%  {\bf s}_{\bf i} \cdot {\bf S}_{\bf j}
  + J_{\rm AF} \sum_{\langle {\bf i,j} \rangle}
  {\bf S}_{\bf i} \cdot {\bf S}_{\bf j}  \nonumber \\
  &+& \lambda \sum_{\bf i}
  (Q_{1{\bf i}}\rho_{\bf i} + Q_{2{\bf i}}\tau_{{\rm x}{\bf i}} 
  +Q_{3{\bf i}}\tau_{{\rm z}{\bf i}})
  +(1/2) \sum_{\bf i} (\Gamma Q_{1{\bf i}}^2
  +Q_{2{\bf i}}^2+Q_{3{\bf i}}^2) + \Delta \sum_{i} S_{\rm iz}^2,
\end{eqnarray}
\end{widetext}
where the hopping amplitude in the ${\bf a}$ direction, $t^{\bf a}_{\gamma \gamma'}$, 
involves the two $e_g$ orbitals that are active for itinerant fermions in manganites, namely $\gamma$ 
and $\gamma^{\prime}$ are $a=x^2-y^2$ or $b=3z^2-r^2$. The
relations among hoppings were extensively discussed before~\cite{dagotto}. Along the $x$-axis we use 
$t^x_{aa}=-\sqrt{3}t^x_{ab}=-\sqrt{3}t^x_{ba}=3t^x_{bb}$, while along the $y$-axis
$t^y_{aa}=\sqrt{3}t^y_{ab}=\sqrt{3}t^y_{ba}=3t^y_{bb}$
Our unit of energy will be $t_{aa}$, which is the same in both directions, 
and the direction supra-index will be omitted for simplicity. 

The angles $\theta$ and $\phi$ correspond to the
orientation of the classical spins at each site, representing the spin of the $t_{2g}$ electrons assumed localized and of magnitude 1. Thus, 
at infinite Hund coupling, the
spin of the mobile $e_g$ quantum electrons and localized $t_{2g}$ spins are perfectly aligned. 
The  lattice degrees of freedom ${Q}$'s are also classical, for simplicity, and they correspond to 
the three Jahn-Teller (JT) modes that are active in manganites~\cite{dagotto}.
$\lambda$ is the dimensionless coupling constant that regulates the strength of the interaction
between electrons and those JT distortions. The fermionic operators
$\tau_{{\rm z}{\bf i}}=\sum_{\sigma}(d_{{\bf i} a\sigma}^{\dag}d_{{\bf i}a\sigma}
-d_{{\bf i} b\sigma}^{\dag}d_{{\bf i}b\sigma})$ are pseudo-spins introduced before~\cite{dagotto}.
The next to last term represents the elastic energy for the distortions, 
with $\Gamma=2$ being the ratio of spring constants for 
breathing- and JT-modes~\cite{sen2012}. Note that, as discussed extensively before, each JT mode is
written explicitly in terms of the oxygens located at the links between Mn atoms. Thus, we are using the proper
$cooperative$ modes. The MC simulation involves those classical oxygen positions. The last term is an anisotropy we introduced
on phenomenological grounds because all materials tend to have a preferred direction for the orientation of the spins. Having this
anisotropy also avoids theorems related to strong fluctuations in rotational invariant systems. We have
explicitly checked that the results barely change if this anisotropy is small and, for simplicity, we used $\Delta = -0.05$ in
all the simulations below. The rest of the notation is standard. 

Note that in several previous studies, to address the rich phase diagrams of manganites with many
antiferromagnetic patterns, the Heisenberg superexchange coupling
$J_{\rm AF}$ between the localized $t_{2g}$ spins at neighboring 
sites ${\bf i}$ and ${\bf j}$ was used. However, here we choose
$J_{\rm AF}=0.00$ because ferromagnetism is the dominant magnetic order in the experiments we are addressing. 

The Monte Carlo (MC) technique used in these simulations for the $t_{2g}$ angles and oxygen displacements has
been extensively discussed before~\cite{dagotto}. The procedure to calculate conductances, and the resistances via their inverses, 
is also standard~\cite{verges00}. In the present effort, we use the cool-down method detailed 
in Ref.~\cite{sen2012} instead of the
standard method where a particular temperature is fixed
and the simulation is run to converge, after thermalization from a random initial state, to a 
particular set of equilibrium configurations. This procedure often fails when the
competing meta-stable states often present are separated by large energy walls, and the Monte Carlo simulations become ``trapped'' in distorted configurations.
Instead, here we start at high temperature with a random
configuration for both the spin and lattice degrees of freedom and anneal slowly
lowering the temperature. This annealing is more costly in time because different temperatures
are not run in parallel, but sequentially instead.

Specifically, we start the cool-down method by selecting a temperature grid. A denser 
temperature grid is used at low temperatures because metastability problems are more likely to arise there.
The temperature grid used for $6 \times 6$ and $9\times 9$ lattices starts at $T=1/3$ and the cooling down involves the temperatures provided 
in~\cite{grid}.
The magnetization calculations for large lattices (up to $15\times 15$) included an additional temperature grid 
that starts at $T=1$ at the beginning of the simulation. This change was needed for better convergence later at lower temperatures, 
since we must use a lower number of MC steps in the largest clusters due to CPU time limitations. 
In our analysis, the simulation starts at a high temperature (again, either $T=1/3$ for 6$\times$6 and 9$\times$9,
or $T=1$ for 12$\times$12 and 15$\times$15) with a random configuration of classical spins and oxygen
lattice degrees of freedom. For the smallest lattice size studied, a $6\times6$ cluster, 
we let the system thermalize for $20,000$ Monte Carlo steps, followed by another $50,000$ to $100,000$ 
steps for measurements, where these measurements are made every 2 steps to decrease autocorrelation
errors. For the $9\times9$ cluster, we use $5,000$ steps for thermalization and $20,000$ for measurements. 

In addition, to increase the statistical quality of the $9\times9$ results, 
we employ the cool-down annealing method 5 independent times, each time using a different random starting 
configuration at high temperatures, and then we average the results.
All these values of the Monte Carlo parameters and convergence process are kept the same for 
all the results discussed below, except for the $12\times12$ and $15\times15$ clusters studied whose specific parameters
are indicated explicitly in the results section. For the 12$\times$12 cluster, 
$5,000$ thermalization and $20,000$ measurement steps were used, while for the $15\times 15$ cluster, 
$1,000$ thermalization and $4,000$ measurement steps were employed.
Because the acceptance ratios of the Monte Carlo simulations often deteriorate 
significantly with decreasing temperature, these acceptance ratios were monitored during the entire 
simulation to ensure that at least they do not fall below 10\% for the lowest temperature considered i.e. $\beta=1/T=300$. 
For details see Ref.~\cite{sen2012}.

%For a fixed value of the electron-lattice coupling $\lambda$, the cool-down process alone might still not be sufficient for full convergence. However, many sets
%of these  model parameters were used in the Monte Carlo runs in parallel (employing several computer nodes)
%and in the end, by mere comparison of energies, it was observed that often for a subset  of those model parameters a convergence 
%to the true ground state was found.
%The existence of a possible new ground state is then confirmed by comparing its energy with those of the neighboring states in the phase 
%diagram. In short, by monitoring the smoothness of the values corresponding to several 
%Monte Carlo observables when  the model parameters are slightly modified, the overall 
%convergence quality of the results can also be monitored.

Once for a fixed parameter set of couplings the true low-temperature ground state is identified  after cooling, the process is then reversed, and this time a ``heat-up'' procedure is carried out, 
namely the simulation starts by using as initial
configuration the properly converged last configuration of the cool-down process at the lowest temperature. 
We use the same thermalization and measurement number of steps in the heat-up simulation as in the cool-down, and 
the same temperature grid, for the $6\times6$ and $9\times9$ clusters. 
Namely in the heat-up process, the study at each temperature starts with
the thermalized configuration of the previous lower temperature.
All these extra steps, involving cooling up and down many times, increase the chances
that truly converged quantities are obtained while improving statistics.

%-----------------------------------------------------------------------------------------
\begin{figure}[H]
\begin{center}
% trim={<left> <lower> <right> <upper>}
 \begin{overpic}[trim = 0cm 0.0cm 0mm 0mm,angle=270,
 width=0.50\textwidth]{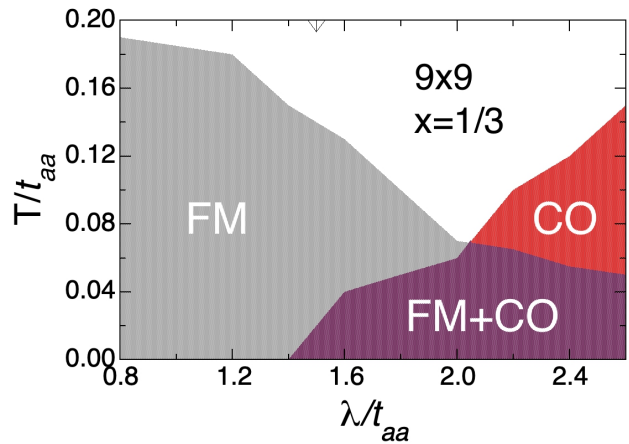}
\end{overpic}
\end{center}
\vspace{-0.6cm}
\caption{Finite temperature Monte Carlo phase diagram of the two-orbital double-exchange model using a
$9\times 9$ lattice at density $x=1/3$, varying the Jahn-Teller strength $\lambda$. All results are in units of the hopping $t_{aa}$.
FM (CO) stands for ferromagnetism (charge order).
As reference, the experimental FM critical temperature of LCMO without strain is $T_C = 250$~K. 
If the left portion of the phase diagram shown is considered to represent LCMO without strain, then the equivalent theoretical
critical temperature is $T/t_{aa} \approx 0.19$. Thus, the insulating critical temperature for high uniaxial strain 150~K found experimentally~\cite{hong2020}
corresponds to $T/t_{aa} \approx 0.11$, placing high-strained LCMO at $\lambda/t_{aa} \approx 2.2$.  }
\label{fig1}
\end{figure}
%-----------------------------------------------------------------------------------------

\section{Results}

We start the description of our results presenting our main conclusion, which is the phase diagram of the double exchange
model previously introduced, where we vary the JT coupling strength and temperature. This main result is shown in Fig.~\ref{fig1}. The phase diagram describes
the evolution of the properties of the model from the metallic charge-disordered
ferromagnetic regime at small $\lambda$ to the insulating charge-ordered ferromagnetic phase at large $\lambda$. These
results are in excellent agreement with those of the recent LCMO membrane experiments where strain effects lead to 
a similar charge-disordered to charge-ordered 
evolution. In our case, strain changes the atomic distances and thus affects the hopping amplitudes, reducing the energy scale
$t_{\rm aa}$ that appears in the denominators of the axes in Fig.~\ref{fig1}. Again, note that $t_{\rm aa}$
is the hopping amplitude between two nearest-neighbor Mn $x^2-y^2$ orbitals (i.e. $a=x^2-y^2$), but strain 
affects all the other hoppings as well. We focus on the hole-doping concentration of $x=1/3$ as in experiments. 

Results were obtained using a 9$\times$9 cluster, but the phase diagram for 6$\times$6 is almost identical (not shown) suggesting
size effects are weak. Our studies on larger clusters up to 15$\times$15 for special cases confirm this assumption. Note that as
the ferromagnetic Curie temperature decreases with increasing $\lambda/t_{\rm aa}$, the charge-ordered critical temperature grows after reaching a critical
value $\lambda/t_{\rm aa} \approx 1.4$.
A region with both FM and CO order exists at robust $\lambda$/$t_{\rm aa}$ and low temperatures. Eventually at large enough $\lambda$/$t_{\rm aa}$, upon 
cooling first a CO phase transition is reached followed later, upon further cooling, by the magnetic transition. 

\begin{figure}
\begin{center}
% trim={<left> <lower> <right> <upper>}
 \begin{overpic}[trim = 0cm 0.0cm 0mm 0mm,angle=0,
 width=0.37\textwidth]{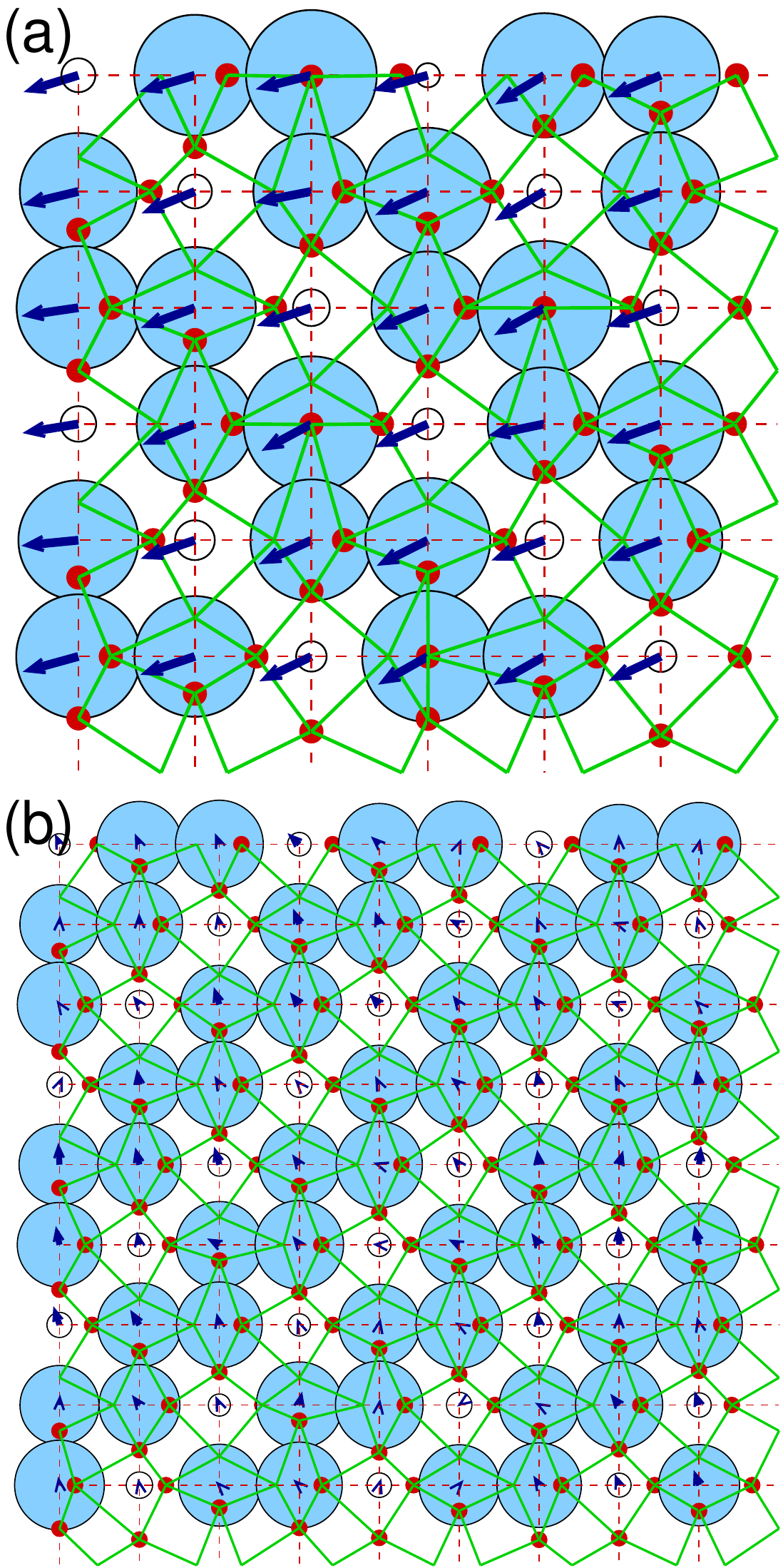}
\end{overpic}
\end{center}
\vspace{-0.6cm}
\caption{Low temperature ($T/t_{\rm aa}$=1/300) typical Monte Carlo equilibrated snapshots of the FM+CO phase 
using (a) $6\times 6$ and (b) $9\times 9$ lattices at density $x=1/3$, after the annealing process described in Sec.~II. Blue circles represent the electronic charge and green lines denote the 
links between oxygen positions which are shown as red dots. 
Dark blue arrows are projections of the classical spins on a two-dimensional surface.}
\label{fig2}
\end{figure}

In the CO phase, the charge configurations that spontaneously emerge from the MC simulations are interesting
and in excellent agreement with {\it ab initio} calculations~\cite{hong2020}. Typical examples for
both 6$\times$6 and 9$\times$9 at very low temperatures are presented in Fig.~\ref{fig2}. They
exhibit diagonal stripes, with the blue circles proportional to the electronic density; thus very small blue circles represent the location of the holes.
The pattern ``two diagonals occupied one empty'' 2-1-2-1 for the diagonal stripes is compatible with a hole density of $x=1/3$. 

%-----------------------------------------------------------------------------------------
%-----------------------------------------------------------------------------------------

We remark that these stripe configurations are in excellent agreement with~\cite{hong2020} and {\it emerge}
in our simulations by starting the MC process at high temperature and slowly cooling down, as described in the previous section. Thus, our results are not
based on variational or mean-field approximations were several candidate configurations are contrasted in energy, but our process assures that this is
the configuration that truly minimizes the energy. Note also that in all cases the spins order ferromagnetically, which is common in
double-exchange models away from $x=0$ in order to take advantage of the large Hund coupling, unless a robust superexchange among the 
$t_{2g}$ spins is introduced~\cite{dagotto}. Scrutinizing with care Fig.~\ref{fig2} readers
can confirm that the spins for the 6$\times$6 cluster -- indicated by arrows -- are mainly parallel to each other although with small oscillations characteristics of an annealing realistic process. The same occurs for the larger system 9$\times$9 albeit with larger angular oscillations. This shows that
the spin plays a secondary role, and the physics is primarily dictated by the well-formed JT distortions -- also shown in Fig.~\ref{fig2} via the oxygens
positions -- that lead to the stripe formation. Finally, we remark again that this stripe configuration was presented years ago~\cite{hotta2001}
when discussing the many possible states that could be present in the rich phase diagrams of manganites under various circumstances. 
In those early calculations a zero-temperature variational relaxation was employed, as opposed to the detailed phase diagram shown here including temperature, due to the limitations associated with available computational power.

%-----------------------------------------------------------------------------------------
\begin{figure}
\begin{center}
% trim={<left> <lower> <right> <upper>}
 \begin{overpic}[trim = 1.5cm 0.0cm 0mm 0mm,angle=0,
 width=0.75\textwidth, height=0.65\textwidth]{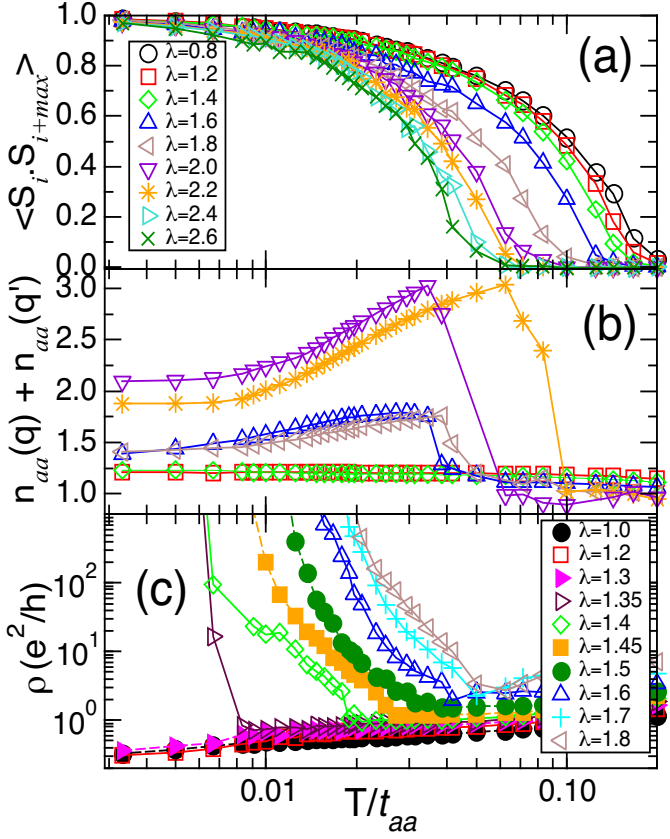}
\end{overpic}
\end{center}
\vspace{-0.6cm}
\caption{(a) Spin-spin correlations at maximum distance, (b) charge structure factors used in the
calculation of the phase diagram of the $9\times 9$ lattice, and (c) resistivity vs. temperature using a $6\times 6$ lattice. The legend in (a) also applies to (b).
}
\label{fig3}
\end{figure}
%-----------------------------------------------------------------------------------------

In order to construct the phase diagram in Fig.~\ref{fig1} using MC simulations, we measured a variety of observables. Consider first Fig.~\ref{fig3}.
In Fig.~\ref{fig3}(a), the spin-spin correlations among the classical spins -- locked parallel to the quantum spins at infinite Hund coupling --
are shown at several couplings $\lambda$. An average over all the sites {\it i} is shown improving the statistics. The ``{\it i+max}'' notation in the vertical axis denotes the site 
furthest away from {\it i}, which on a 9$\times$9 cluster means Euclidean distance $4\sqrt{2}$ [example sites (1,1) and (5,5)]. Clearly, these spins are
ferromagnetically correlated. Figure~\ref{fig3}(a) shows that this correlation decreases with increasing temperature up to a fairly well-defined temperature.
As shown later in our analysis, the best estimation for the Curie temperature is obtained from extrapolation of the results with negative curvature -- as
expected in the bulk -- and neglecting the narrow portion with positive curvature which is a size effect. But even without accounting for this detail, 
qualitatively the results are clear: with increasing $\lambda$ the FM Curie temperature rapidly decreases (although it remains finite in the range
shown in Fig.~\ref{fig1}).

In Fig.~\ref{fig3}(b) the charge structure factor $n_{\gamma\gamma'}(\textbf{q})$ is shown. This quantity was 
obtained by Fourier transforming the quantum charge correlations 
$\< n_{\gamma}n_{\gamma'}(\textbf{r}) \> \equiv \<n_{i\gamma}n_{j\gamma'}\>$ 
among the $e_g$-orbitals $a=x^2-y^2$ and $b=3z^2-r^2$, in all combinations $aa$, $bb$, $ab$, and $ba$, 
and averaging over MC steps during measurements. These charge correlations are given as:

\begin{eqnarray}
\<n_{i\gamma}n_{j\gamma'}\> = \< d_{i\gamma}^\dag d_{i\gamma}d_{j\gamma'}^\dag d_{j\gamma'}\> 
&=& (1-G_{ii})(1-G_{jj}) \nonumber \\  &-& G_{ij}G_{ji},
\end{eqnarray}

\noindent where the Green's function $G_{ij}$ is defined as:

\begin{eqnarray}
G_{ij} \equiv \sum_{\tau}^{2N}U_{i\tau}\frac{1}{1+e^{-\beta\epsilon_{\tau}}}U_{j\tau}^{\dag}.
\end{eqnarray}

\noindent Here $U$ is the unitary matrix that diagonalizes the fermionic
Hamiltonian with eigenvalues $\epsilon_{\tau}$. The rest of the notation is standard and the spin index in the 
definition of the charge correlations is suppressed for clarity. 
In Fig.~\ref{fig3}(b) we only show $n_{aa}$ for simplicity because it is the dominant one, 
but $n_{bb}$ is similar in magnitude, while the non-diagonal $ab$ and $ba$ are weaker.
The wave-vectors corresponding to the two possible diagonal stripes are defined as $\textbf{q} = (2\pi/3,4\pi/3)$ and 
$\textbf{q$^\prime$} = (4\pi/3,2\pi/3)$. Note that in the figures we show the total charge structure factor adding both stripes wavevectors
$n_{aa}(\textbf{q})+n_{aa}(\textbf{q$^\prime$})$ because different values of the electron-phonon coupling $\lambda$ can converge
to either one or the other of these stripes (i.e. in Fig.~\ref{fig2}(a) we show a right-to-left diagonal stripe while
Fig.~\ref{fig2}(b) shows a left-to-right diagonal stripe). Plotting the total structure factor adding the two momenta 
allows us to avoid having to track the orientation of the diagonal stripes for each $\lambda$.

Returning to the charge correlations Fig.~\ref{fig3}(b) vs. temperature, they always have a nonzero value due to short distance
effects, but cases such as $\lambda=1.2$ (red symbols) show the charge correlation is totally flat and weakly varying with temperature indicating 
that at this electron-lattice coupling there is no action in the charge ordering channel. However, at $\lambda \approx 1.6$ 
these correlations develop a non-trivial structure increasing with temperature, and then
suddently dropping. This  provides an indication of the CO critical temperature $T_{CO}$ which grows with increasing $\lambda$ in the FM+CO phase, 
as expected. Finally,
in Fig.~\ref{fig3}(c) the resistivity vs. temperature is shown. In this quantity, a clear transition from a metal at high temperature to an insulator
at low temperature is found, and the temperature location of this MIT is in good agreement with the results of Fig.~\ref{fig3}(b), providing a consistency
test to our procedure. Moreover, the shape of the curves with a CO ground state resemble the experimental result~\cite{hong2020} at high uniaxial strain 8\%.

%-----------------------------------------------------------------------------------------
\begin{figure}[H]
\begin{center}
% trim={<left> <lower> <right> <upper>}
 \begin{overpic}[trim = 0cm 0.0cm 0mm 0mm,angle=0,
 width=0.55\textwidth]{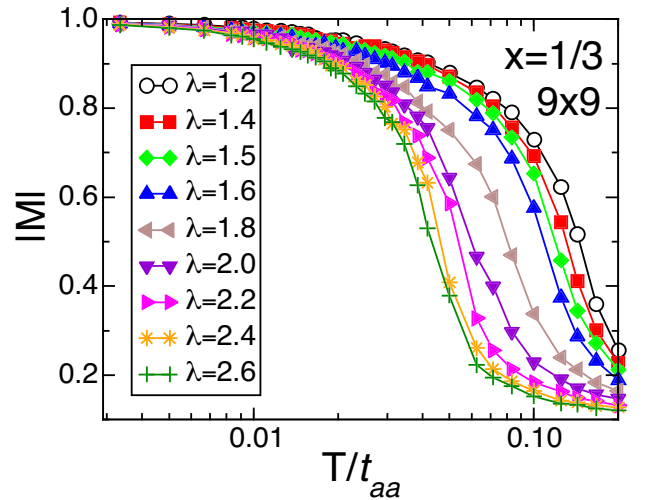}
\end{overpic}
\end{center}
\vspace{-0.6cm}
\caption{ Magnetization (in absolute value, see text) as a function of temperature using a $9\times 9$ lattice, parametrized with the JT coupling $\lambda$. }
\label{fig4}
\end{figure}
%-----------------------------------------------------------------------------------------

In Fig.~\ref{fig4}, the absolute value of the magnetization is shown. This quantity is defined as
\begin{eqnarray}
|{\rm M}| = \frac{1}{N}\sqrt{{\bf M}.{\bf M}},
\end{eqnarray}
\noindent where the vector total magnetization involving the classical spins is
\begin{eqnarray}
{\bf M}= \sum_i {\bf S}_i,
\end{eqnarray}
and $N$ is the total number of sites.
Qualitatively $|M|$ should behave similarly as the correlations in Fig.~\ref{fig3}(a). Indeed Fig.~\ref{fig4} has
the characteristic behavior of an order parameter normalized to 1 at low temperature. Because in this magnetization the
short-distance spin-spin correlations are included, the region with positive curvature near $T_C$ is larger than in Fig.~\ref{fig3}(a)
but qualitatively the results are very similar and the curves smoothly decrease towards lower temperatures as $\lambda$ increases, as expected. 
Thus, either measuring the magnetization or the largest distance spin-spin correlations similar results are obtained.

%-----------------------------------------------------------------------------------------
\begin{figure}[H]
\begin{center}
% trim={<left> <lower> <right> <upper>}
 \begin{overpic}[trim = 0cm 0.0cm 0mm 0mm,angle=270,
 width=0.5\textwidth]{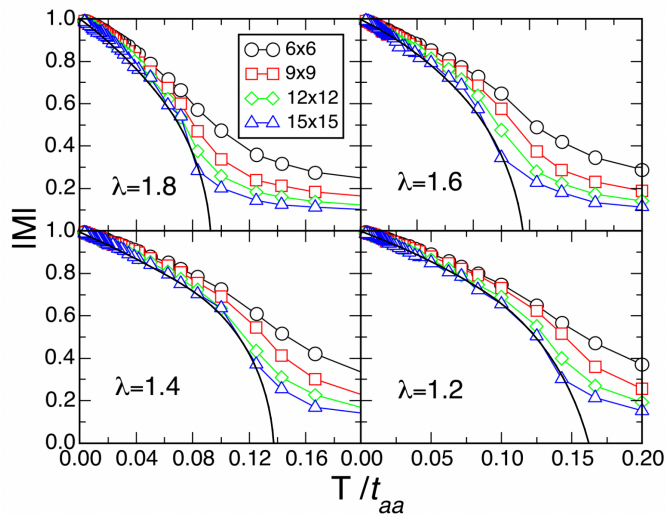}
\end{overpic}
\end{center}
\vspace{-0.6cm}
\caption{Magnetization (absolute value) vs. temperature for lattice sizes up to $15\times 15$ at $x=1/3$ density and the values of $\lambda$ indicated. 
The black lines without points are extrapolations using the portion with negative curvature of the 15$\times$15 cluster curve. } 
\label{fig5}
\end{figure}
%-----------------------------------------------------------------------------------------

Finally, Fig.~\ref{fig5} contains a finite-size analysis of our results in the middle region between the FM and FM+CO phases.
Shown are results for several cluster sizes. A similar careful analysis could not be carried out for all observables because of the limited
number of MC steps allowed within a reasonable amount of CPU time. In particular, measurements of resistivity and charge-charge correlations tend
to be noisier, thus in those cases consistency between 6$\times$6 and 9$\times$9 was considered sufficient. But for the relatively simpler magnetic properties
we could extend the analysis to 15$\times$15 clusters. As Fig.~\ref{fig5} shows, the results are smooth varying cluster size and the region of positive curvature
due to size effects decreases, allowing for the smooth extrapolation to the bulk limit shown where the resulting curve has the negative curvature in all the
ordered range as expected.

\section{Conclusions}

We have theoretically studied the manganite LCMO in the regime between small and large Jahn-Teller coupling to address recent exciting 
experiments~\cite{hong2020} where extreme tensile strain was applied
to LCMO membranes beyond 8\% uniaxially and 5\% biaxially, a high strain never reported before. These values are 
well beyond those achieved in standard superstructures
where strain is controlled by growth on a discrete set of available 
substrates and it is affected by strain-driven defect formation. 
Transport measurements in the novel LCMO highly-strained membranes indicated a transition from a metallic ground state at low strain to an exotic insulating
state at high strain. Early theory efforts several years ago already predicted what the ground state in the regime of a large Jahn-Teller coupling should be,
and it is in excellent agreement with DFT calculations~\cite{hong2020}.
Large Jahn-Teller coupling is effectively produced by high strain since a bandwidth $W$ decreasing effectively leads to a Jahn-Teller coupling $\lambda$
increasing because what controls the true strength is the ratio $\lambda/W$.

In our effort incorporating temperature effects by a careful Monte Carlo annealing process from high temperature, we showed that the insulating state is indeed
the exotic ground state predicted years ago at zero temperature. This state displays diagonal stripes of charge at hole doping $x=1/3$, 
with regions in between the doped-rich stripes containing orbital order in a staggered pattern resembling the spin staggeredness 
in the more widely discussed stripes in cuprates and nickelates.
Our research places the existence of these unexpected complex ground states on robust grounds, extends to cover the entire temperature range, and provides
the complete phase diagram. Moreover, after the $x=1/3$ state is confirmed, then all the early predictions for stripes at $x=1/m$ ($m$ integer) 
receive indirect validation as ground states. Confirming this broad family of stripes at hole densities different from $x=1/3$ 
should be a possible direction of research for the exciting experimental new procedure that creates LCMO membranes at high strain.

\section{Acknowledgments} 

C.~\c{S}. acknowledges the resources provided by the Advanced Computational Facility (ACF) of 
the National Institute for Computational Sciences (NICS). 
E.~D. was supported by the U.S. Department of Energy (DOE),
Office of Science, Basic Energy Sciences (BES), Materials Science and
Engineering Division.

%%%%=========================================================%%%
%\bibliographystyle{naturemag}
%\bibliography{citations}

\begin{references}
\bibitem{grid} For the 6$\times$6 and 9$\times$9 clusters the specific temperature grid used started at $\beta=1/T=3$ and then it involved
$\beta = 1/T= 5,~6,~7,~8,~10,~12,~14,~16,~20,~24,~26,~29,~32,~34$, $37,~40,~43,~48,~51,~53,~56,~60,~64,~68,~73,~80,~90,~100,~110,$ $120,~150,~200,~300$.
\end{references}

\end{document}